\documentclass{tufte-handout}

\usepackage{enumitem}
\usepackage{amsmath}

\usepackage{graphicx}
\setkeys{Gin}{width=\linewidth,totalheight=\textheight,keepaspectratio}
\graphicspath{{graphics/}}

\usepackage{booktabs}

\usepackage{units}

\usepackage{fancyvrb}
\fvset{fontsize=\normalsize}

\usepackage{multicol}

\usepackage{lipsum}

\usepackage{hyperref} 
\usepackage{enumitem}
\setlist{nosep}

\title{A Preliminary Model for the Design of Music Visualizations \thanks{This paper appears as a poster at PacificVis 2021}}
\author{Swaroop Panda, Shatarupa Thakurta Roy}
\date{Indian Institute of Technology Kanpur}  
\begin{document}

\maketitle

\begin{abstract}
Music Visualization is basically the transformation of data from the aural to the visual space. There are a variety of music visualizations, across applications, present on the web. Models of Visualization include conceptual frameworks helpful for designing, understanding and making sense of visualizations. In this paper, we propose a preliminary model for Music Visualization. We build the model by using two conceptual pivots – Visualization Stimulus and Data Property. To demonstrate the utility of the model we deconstruct and design visualizations with toy examples using the model and finally conclude by proposing further applications of and future work on our proposed model.
\end{abstract}

\section{Introduction}

Music is primarily aural in nature. Music is recorded, processed and distributed in a variety of data formats; such as digital audio, MIDI (Musical Instrument Digital Interface), and other symbolic structures like sheet notations. These data formats (or representations) contain rich information about the music file and thus is available for the purpose of analysis.

Visualization of music refers to the transformation of the data from the aural to the visual space. Music is made visible. There are a plenty of existing music visualizations across the web. They are found in a variety of formats. Their wide range of applications makes them complex and hard to decipher. These visualizations range from ordinary spectrograms (used for sophisticated Machine Learning tasks) to fancy graphics in media players and streaming services accompanying the music (for a augmented and a pleasing user experience) to fanciful tree maps representing a playlist (used in exploratory music databases). Foote in \cite{foote1999visualizing} visualizes digital audio across time using acoustic similarity between any two instants of the audio clip using a 2 dimensional representation. Pampalk et al. \cite{pampalk2001islands} builds Islands of music for a Graphical User Interface of Music Archives using Self-Organising Maps. Hiraga \cite{hiraga2002music} uses music visualization for learning. Tzanetakis \cite{cooper2006visualization} lists and elaborates on the use of visualization in the Music Information Retrieval (a subset of the larger music research) community. 
Morchen et al. in \cite{morchen2005databionic} clusters and visualizes music collections by using features that capture the idea of perceptual similarity. As evident from the cited works ,they have different uses, different contexts and different scope. The demystification of such a large and diverse collection of music visualizations would be useful for building a set of guidelines or a model for music visualization designers.    


A visualization model captures the latent processes within the visualization design process. These models aid the visualization designer in the development process; by providing a basic framework or a set of guidelines. These models are also helpful to understand, make sense of and appreciate visualizations.  Munzner \cite{munzner2009nested} provides a nested model for visualization design and validation. Liu et al. \cite{liu2008distributed} provides a distributed cognition theoretical framework for understanding and appreciating visualizations. Similarly,  Kindlmann et.al. \cite{kindlmann2014algebraic} provides an algebraic process for visualization design. 

In this work, we present such a preliminary model for music visualization. We build the model by analysing properties of the digital audio data and the subsequent visual transformation. We demonstrate a two-fold validation of the model- by deconstruction and by design. We conclude by discussing potential further improvements and future work on this model.

\section{The Model}
We begin constructing the model by exploring the purpose of music visualization. What is the motivation to visualize music? For the sake of nomenclature, we bisect the purpose of music visualization into two components: \textit{Functional} and \textit{Aesthetic}. The Functional fork contains visualizations intended to meet well defined and bounded ends; recommendation, production or music exploration. The purpose of the music visualization is to demonstrate or exhibit a well-defined system, such as a digital workstation. The DAW is a visualization intended to provide the producer with tools to execute musical ideas. The Aesthetic fork contains visualizations intended to meet more open and unbounded ends; such as sensory augmentation or enhancing overall music experience. An example of this is the visualizations present in music players like VLC, where the intent of the visualization is to enhance the user experience of a music track by adding pleasing visual forms. Another differentiator between the functional and aesthetic forks is that for functional visualizations there are rules and guidelines to follow. The DAW has in-built signal processing algorithms which require certain guidelines. The aesthetic fork gives much more liberty to the designer for creating the visualizations.

\begin{figure}[h!]
 \centerline{
 \includegraphics[width=9cm]{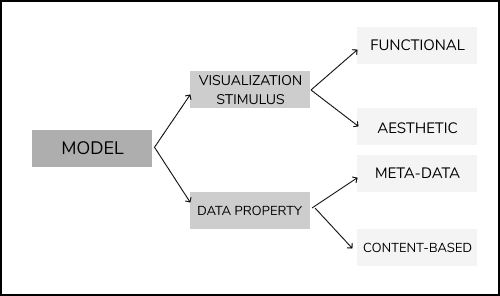}}
 \caption{The model with the conceptual pivots.}
 \label{fig:tm8}
\end{figure}
The other crucial aspect in Music Visualization is to realize what exactly is being visualized; the \textit{meta-data} of the audio file or some \textit{content-based} extracted from the audio file. Meta-data are the annotations associated with the audio file. They are those properties that are not extracted from the audio content but are labels provided by experts (or the artists themselves). These meta-data can include song name, artist name, genres, date of release or albums. Content-based properties, on the other hand, are extracted from the audio files from the audio content. These properties are usually extracted by means of an algorithm. An instance of such a content-based visualization is a spectrogram. A spectrogram is obtained by using a Fourier transform algorithm on the audio data. It is thus a property extracted from the audio file.

Meta-data is data present in a variety of forms across taxonomies. Dates of release are a set of dates, Artist names are sets of strings, Billboard ratings are a bunch of numbers and so on. Visualizing meta-data then involves visualizing these data taxonomies \cite{shneiderman1996eyes}; using the existing or new visualizations. It is noted that these taxonomies are not independent but are associated with audio files. Content-based visualizations are computational in nature and have guidelines based on their governing algorithms. A spectrogram has well defined x-y axis and colours present have specific meanings attached to them. The data type and taxonomy depends on the algorithms used to extract the properties of the audio file.

These two pivots are not independent, but are associated in complex ways. Content-based data is mostly used for a functional use (like spectrograms for music artist classification) while meta-data is used for aesthetic purposes (like a pleasing layout of a song collection). However, spectrograms have been used for a pleasing experience and meta-data have been used for automatic data-annotation.

\section{Model Validation}
The music visualization model, we suggested, is helpful for aiding designers in the development process and for the purpose of understanding and making sense of music visualizations. We validate the model in a two-fold manner; we use the model to deconstruct exiting visualizations and we use the model to design visualization prototypes. 
For deconstruction, as a toy example, we use a self-similarity matrix. A self-similarity matrix for an audio is a table defining how similar are the small discrete elements of the audio file with respect to each other. The x-axis of this matrix contains the small piece-wise components of the audio in the correct order. Similarly, the y-axis too contains the piece-wise components in the order. The self-similarity matrix is obtained from the content of the audio file using an algorithm. It is thus, extracted from the content. This self-similarity matrix is majorly used for functional purposes. The work in \cite{silva2018fast} has used a similarity matrix for cover-song recognition. 

We now demonstrate the utility of our model by designing two visualizations. This model, though does not propose a full-fledged workflow, acts as a guideline for the visualization design process. We show such two examples, trying to capture all all the forks of our model.

Say the design problem is to display a playlist of a selection of songs from the Beatles for a streaming service. We proceed with the design in the following manner. The playlist is arranged chronologically along a timeline, grouped by the corresponding album. This is an instance of \textit{meta-data} visualization. The song name and album listing are properties that are not derived from the content. This visualization is \textit{aesthetic} in essence. The visual layout of the playlist isn't adhering to any technical guidelines, but rather is to provide a pleasing experience to the user to navigate through the Beatles’ songs. 

Another design problem is to provide for a music equalizer for a streaming service. This equalizer is supposed to independently control the Harmonic and Percussive components of an audio file in real-time. The harmonic component of an audio file contain the pitched instruments (piano, guitar, human voice) while the percussive components comprise of non-pitched instruments (drums). This is a strictly \textit{content-based} visualization. The harmonic and percussive components are extracted from the audio clip by means of an algorithm. This visualization serves as a \textit{functional} use. The user can choose to tune the equalizer for each of these components.

\begin{figure}[htb]
 \centerline{\framebox{
 \includegraphics[width=9.5cm]{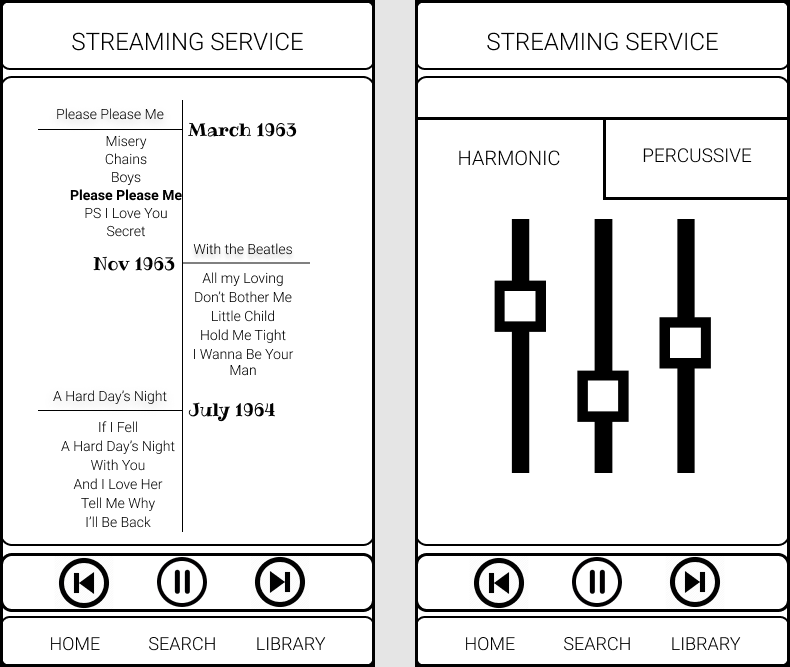}}}
 \caption{Sample Prototypes of Music Visualizations.}
 \label{fig:tm183}
\end{figure}

\section{Discussion \& Conclusion}
Further this model can be tuned into more specific tasks; for instance functional can be forked into recommendation, streaming, surfing. Also this model can accommodate the other music data types such as notations or MIDI data. Each of these data types encapsulate peculiar music features; which would be interesting to visualize. A design workflow, describing a complete Visualization Design process, can be developed from this model. This would encourage more designers to develop music visualizations. The model was primarily built to facilitate and help designers to develop new music visualizations. So, an important component of validation of this model includes evaluation conducted with the designers. This involves conducting carefully crafted long term studies to ascertain the impact of this preliminary model.

To reiterate, in this paper we proposed a preliminary model for music visualization.  Our model is based on two conceptual pivots. Both of these conceptual pivots are further bifurcated into Functional - Aesthetic and Meta Data - Strictly Content based forks respectively. We validated such a framework by looking at an existing visualization as well as designing two prototypes. We concluded by discussing and proposing further improvements on this model. 

\bibliography{sample-handout}

\begin{thebibliography}{10}
\providecommand{\natexlab}[1]{#1}
\providecommand{\url}[1]{\texttt{#1}}
\expandafter\ifx\csname urlstyle\endcsname\relax
  \providecommand{\doi}[1]{doi: #1}\else
  \providecommand{\doi}{doi: \begingroup \urlstyle{rm}\Url}\fi

\bibitem[Cooper et~al.(2006)Cooper, Foote, Pampalk, and
  Tzanetakis]{cooper2006visualization}
Matthew Cooper, Jonathan Foote, Elias Pampalk, and George Tzanetakis.
\newblock Visualization in audio-based music information retrieval.
\newblock \emph{Computer Music Journal}, 30\penalty0 (2):\penalty0 42--62,
  2006.

\bibitem[Foote(1999)]{foote1999visualizing}
Jonathan Foote.
\newblock Visualizing music and audio using self-similarity.
\newblock In \emph{ACM Multimedia (1)}, pages 77--80, 1999.

\bibitem[Hiraga et~al.(2002)Hiraga, Watanabe, and Fujishiro]{hiraga2002music}
Rumi Hiraga, Fumiko Watanabe, and Issei Fujishiro.
\newblock Music learning through visualization.
\newblock In \emph{Second International Conference on Web Delivering of Music,
  2002. WEDELMUSIC 2002. Proceedings.}, pages 101--108. IEEE, 2002.

\bibitem[Kindlmann and Scheidegger(2014)]{kindlmann2014algebraic}
Gordon Kindlmann and Carlos Scheidegger.
\newblock An algebraic process for visualization design.
\newblock \emph{IEEE transactions on visualization and computer graphics},
  20\penalty0 (12):\penalty0 2181--2190, 2014.

\bibitem[Liu et~al.(2008)Liu, Nersessian, and Stasko]{liu2008distributed}
Zhicheng Liu, Nancy Nersessian, and John Stasko.
\newblock Distributed cognition as a theoretical framework for information
  visualization.
\newblock \emph{IEEE transactions on visualization and computer graphics},
  14\penalty0 (6):\penalty0 1173--1180, 2008.

\bibitem[M{\"o}rchen et~al.(2005)M{\"o}rchen, Ultsch, N{\"o}cker, and
  Stamm]{morchen2005databionic}
Fabian M{\"o}rchen, Alfred Ultsch, Mario N{\"o}cker, and Christian Stamm.
\newblock Databionic visualization of music collections according to perceptual
  distance.
\newblock In \emph{ISMIR}, pages 396--403, 2005.

\bibitem[Munzner(2009)]{munzner2009nested}
Tamara Munzner.
\newblock A nested model for visualization design and validation.
\newblock \emph{IEEE transactions on visualization and computer graphics},
  15\penalty0 (6):\penalty0 921--928, 2009.

\bibitem[Pampalk(2001)]{pampalk2001islands}
Elias Pampalk.
\newblock \emph{Islands of music: Analysis, organization, and visualization of
  music archives}.
\newblock na, 2001.

\bibitem[Shneiderman(1996)]{shneiderman1996eyes}
Ben Shneiderman.
\newblock The eyes have it: A task by data type taxonomy for information
  visualizations.
\newblock In \emph{Proceedings 1996 IEEE symposium on visual languages}, pages
  336--343. IEEE, 1996.

\bibitem[Silva et~al.(2018)Silva, Yeh, Zhu, Batista, and Keogh]{silva2018fast}
Diego~F Silva, Chin-Chia~M Yeh, Yan Zhu, Gustavo~EAPA Batista, and Eamonn
  Keogh.
\newblock Fast similarity matrix profile for music analysis and exploration.
\newblock \emph{IEEE Transactions on Multimedia}, 21\penalty0 (1):\penalty0
  29--38, 2018.

\end{thebibliography}
\bibliographystyle{plainnat}

\end{document}